\documentstyle[multicol,epsf,aps]{revtex}
\begin{document}
\def\n{\eta}
\def\t{\tau} 
\def\D{\Delta}
\def\r{\rho}
\def\p{\psi}
\def\s{\sigma}
\def\g{\gamma}
\def\st{\sigma_{stick}}
\def\sl{\sigma_{slip}}
\def\ra{\rightarrow}
\def\S{\sigma}
\def\sb{\bar\sigma}
\def\om{\omega}
\def\l{\lambda}
\def\be{\beta}
\title{Modeling eruptions of Karymsky volcano}

\author{A.~Ozerov$^*$,
I.~Ispolatov$^{\dagger}$, and J.~Lees$^\P$}
\address
{$^*$\ldots Russian Academy of Science, Petropavlovsk-Kamchatsky, Russia,
$^{\dagger}$Center for Studies in Physics and Biology, Rockefeller University, 
New York, NY 10021,
$^{\P}$Department of Geological Sciences, University of North Carolina
Chapel Hill, NC 27599-3315.
} 

\maketitle

\begin{abstract} 

\noindent  
A model is proposed to explain temporal patterns of activity in a class 
of periodically  exploding Strombolian-type volcanos. 
These patterns include major events (explosions) which follow each other every
10-30 minutes and subsequent tremor with a typical period of 1 second. This
two-periodic activity is thought to be caused by two distinct mechanisms 
of accumulation of the elastic energy in the moving magma column:
compressibility of the magma in the lower conduit and viscoelastic response of the almost solid magma plug on the top. A release of the elastic energy 
happens when a stick-slip dynamic phase transition in a  
boundary layer along the 
walls of the conduit occurs; this phase transition is driven by the 
shear stress accumulated in the boundary layer. 
The first-order character and intrinsic 
hysteresis 
of this phase transition explains the long periods of inactivity in the 
explosion cycle.
Temporal characteristics of the model are found to be qualitatively similar to
the acoustic and seismic signals recorded at Karymsky volcano in Kamchatka. 


\end{abstract}

\section{Introduction}
A wide variety of types of volcanic activity exists: from devastating 
explosions separated by calm periods of many years or even centuries 
to a steady continuous outpouring of magma. In this paper we study 
mechanisms that give rise to a Strombolian-type volcanic activity, which is somewhere between these
two extremes and 
characterized 
by regularly repeated explosions (10-400 per day) followed  by relatively calm periods. 
Often in the course of a longer explosion,
the gas and ash emission exhibits audible and visible modulations
 with a rather robust period of about 1 second. 
These modulations,  often found on the seismograms
of Strombolian-type eruptions, received a special name,
``chugging'' (Benoit and McNutt, 1997), because they resemble 
a periodic noise produced by a steam engine.
An example of a volcano exhibiting such activity, we consider 
Karymsky volcano on Kamchatka peninsula in the Far East of Russia.

We believe that these Strombolian-type eruption patterns are caused by 
peculiarities of the motion of the magma column,
and  propose a 
model that describes the magma motion in a volcanic conduit 
as a creep flow of viscoelastic compressible 
medium with shear-stress-dependent boundary conditions. 
We show that the dynamics of the model
indeed exhibits 2 levels of quasiperiodic behavior and resembles the eruption
patterns of the Strombolian-type volcanos
both on long and short time scales. The physical transparency of the
suggested model allows us to express
the observed time scales using the material properties of the magma. 

Our paper is organized in the following way:
first, we briefly describe geological aspects of the current 
eruption of Karymsky volcano. Then, after referring to the existing 
explanations 
of Strombolian-type activity, the model is
formally introduced. We derive simple analytical expressions
for a dormant time, a duration of explosion , and a typical period
of tremor, and then present a numerical solution to a system
of dynamical equations describing the motion of the magma column.
The paper is concluded with a discussion of results and possible directions 
for further studies.

\section{The Karymsky Eruption of 1996 - 2000}
\begin{figure}
\centerline{\epsfxsize=7cm \epsfbox{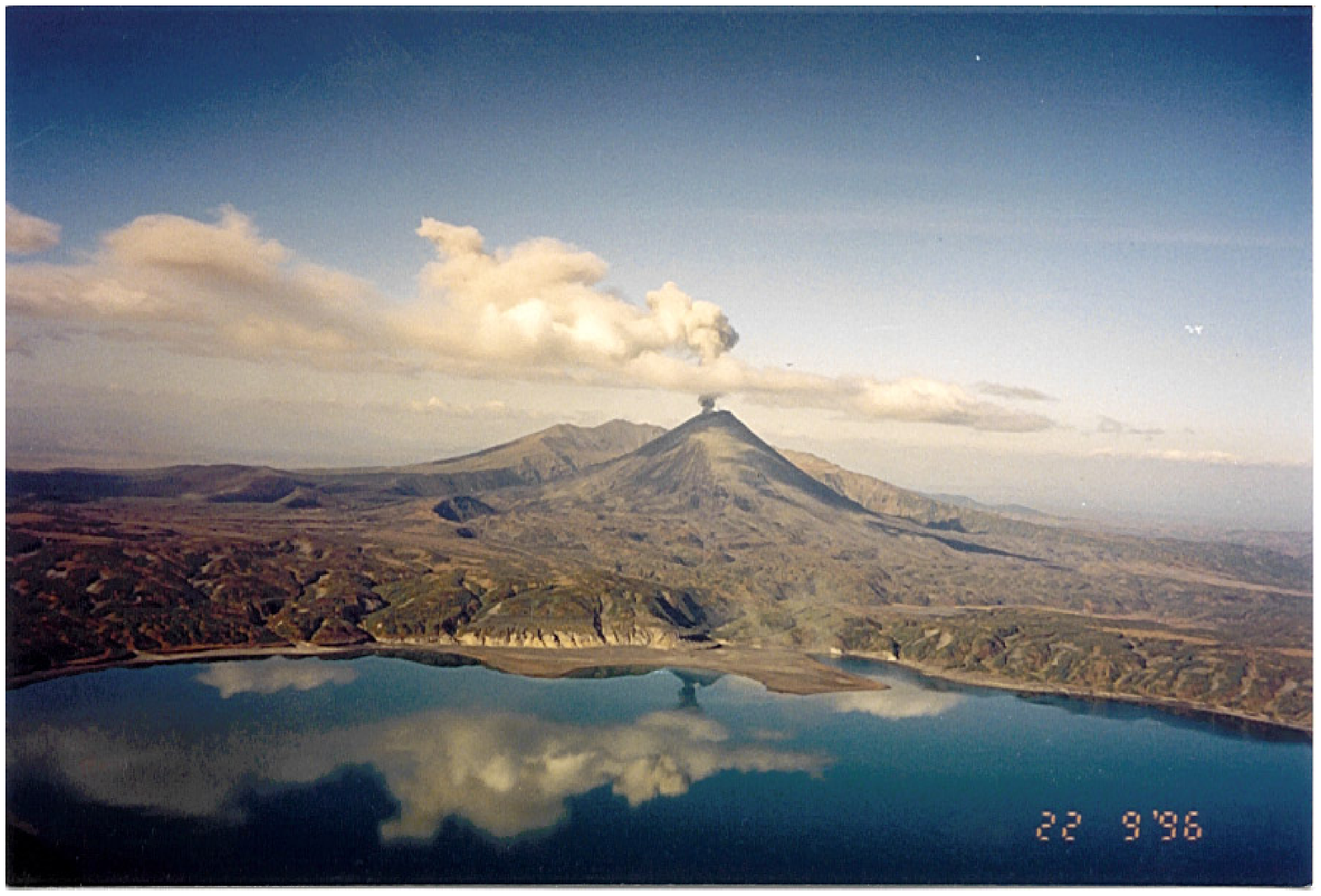}}
\noindent
{\small {\bf Fig.~1.}
Karymsky volcano.
}
\end{figure}

Karymsky volcano (Fig.~1), located in the central part of the 
East Kamchatka volcanic belt,
is one of the most active volcanoes in the far-eastern Russia. 
It is a typical andesite stratovolcano 
composed of a $7700-7800$-year
old caldera $\sim 5$~km in diameter (Ivanov et al., 1991) and a more recent 
cone which started
growing about $5300$ years ago and at present (September 2000) is $\sim 700$~m high.
The cone is composed of accumulated lava and pyroclastic materials, and the 
summit of the
cone is capped by a double crater. The active part of the crater,
formed during the most recent eruption, widened
from 90~m in 1996 to 190~m in 2000 (Fig.~2). The absolute height of the 
volcano is 1549~m.
\begin{figure}
\centerline{\epsfxsize=7cm \epsfbox{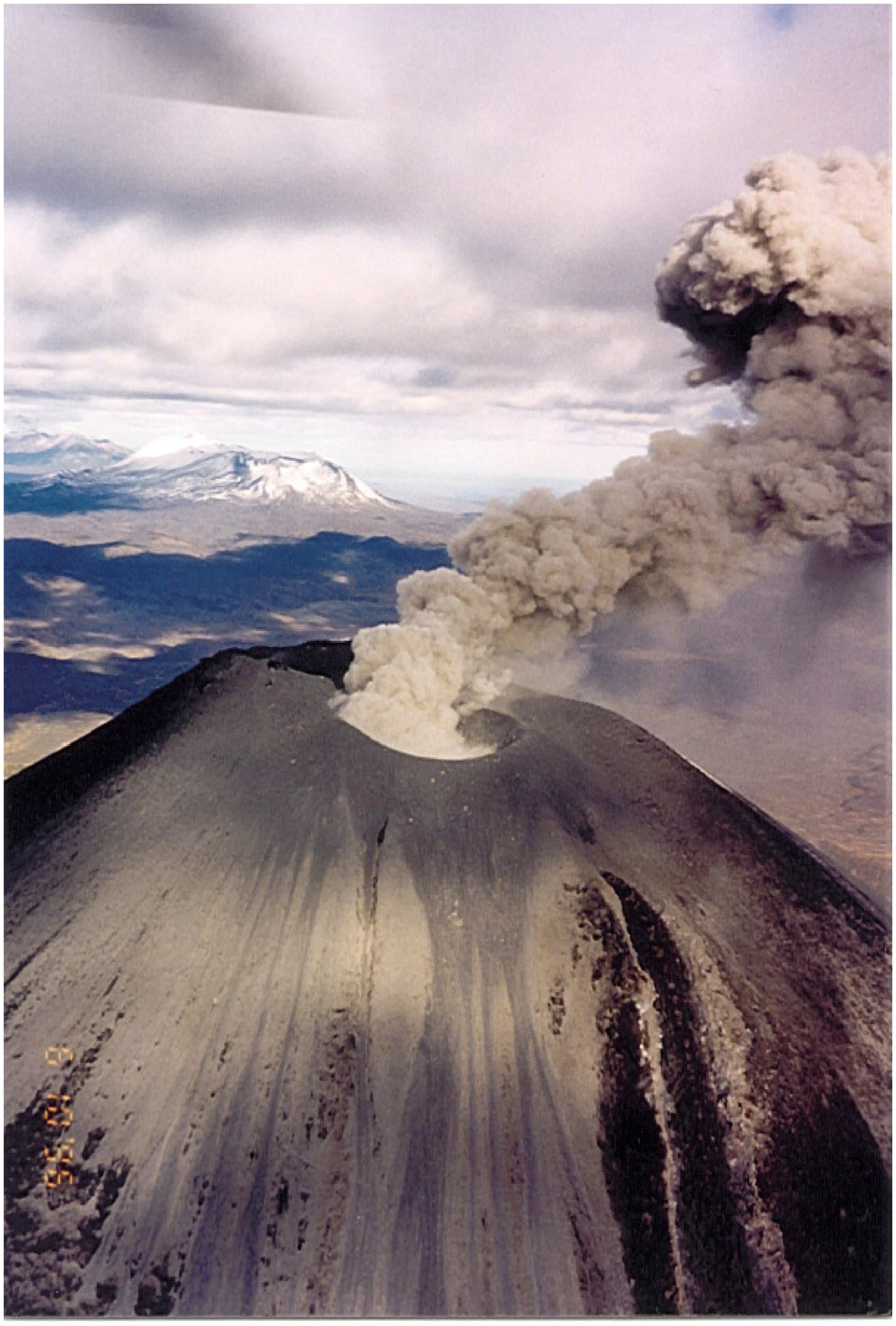}}
\noindent
{\small {\bf Fig.~2.}
Double summit crater at Karymsky volcano.
}
\end{figure}
Karymsky activity usually consists of
long effusive-explosive eruptions, with the eruption previous to the one 
being studied taking place
in 1970-1982.
The current eruption at Karymsky volcano began in January 1996 
and apparently entered a lull phase in December 2000.
Each eruption consists mostly of the discrete 
quasiperiodic gas and ash bursts sending plumes 100-1000~m 
above the crater rim (Fig.~3).
\begin{figure}
\centerline{\epsfxsize=7cm \epsfbox{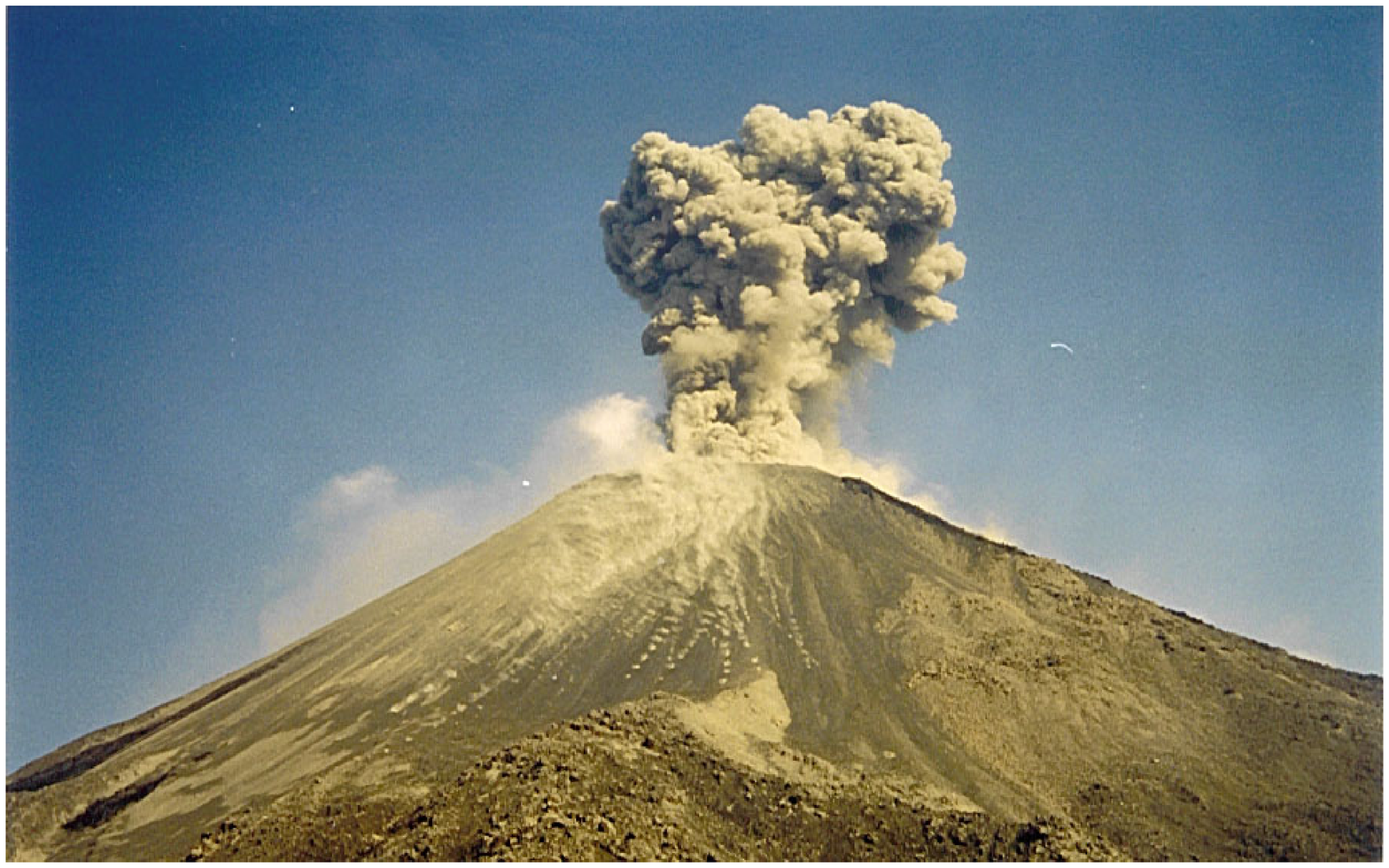}}
\noindent
{\small {\bf Fig.~2.}
Gas and ash plums at Karymsky volcano.
}
\end{figure}
Ash and steam clouds have been observed extending downwind for
10-50~km, rarely 100-200~km.

A lava flow field, produced by the very viscous 
magma, runs down the
south-western slope of the cone and reaches the length
of $\sim 1.4$~km and the width $\sim 100-200$~m. 
Even during the periods of highest activity,
the motion of the lava flows was
very slow (few meters per day) and was noticeable only by
periodic rock-falls, happening every 5-15 minutes.

During particularly strong explosions,
the eruptive column collapsed causing
pyroclastic flows. 
In the lower part of the eruptive column, 
hot volcanic bombs of the sizes up to 2, rarely 5~m, 
were frequently observed. The bombs were dense and often covered
with a bread-crust-like structure formed during cooling.
On the surface of some bombs, noticeable
extrusive tracks were found; they were scratched on the very viscous 
moving magma by the protruding parts of the magma channel.

Total volumes of pyroclastic and effusive materials deposited on the cone
of Karymsky during the eruption of 1996-2000 is estimated to be
0.0274 and 0.0229 km$^3$, respectively.
Eruptive products, which include magma, volcanic bombs, and ash, 
predominantly consist of andesite, usually of black 
and dark grey color. The most common mineral in the eruptive products is 
plagioclase (20-25~\%). 
Crystals of plagioclase  can reach a size 
of 3-5~mm. The concentration of olivine and pyroxene is less than 1\%, 
their crystal 
sizes are usually 2-3~mm. A chemical content of the andesite, 
averaged over 6 samples from the 1996 - 2000 eruption, 
is listed below:
$SiO_2$ -- 61.54, $TiO_2$ -- 0.86, $Al_2 O_3$ -- 16.67, $Fe_2 O_3$ -- 2.43,
$FeO$ -- 5.12, $MnO$ -- 0.11, $MgO$ -- 1.99, $CaO$ -- 5.31, $Na_2 O$ -- 3.69,
$K_2 O$ -- 1.58, $P_2 O_5$ -- 0.25.

The feeding system of 
Karymsky volcano was studied by various methods, including
gravity surveys, aeromagnetic surveys, photogrammetry, seismology and
geodesy (Zubin, 1971; Maguskin, 1982; Shirokov, 1988). 
Despite slight discrepancies in the estimates
of the depth of the upper border of magma chamber and its size, 
these data agree
that underneath the Karymsky volcano in the close proximity to the surface 
there exists a spherical magma chamber. The upper border of this 
chamber is in few kilometers below the sea level. The diameter of the 
magma chamber is estimated to be between 1.5 and 7~km. A magma channel 
of a diameter of 100-200~m
leads from the magma chamber to the top crater.

\section{dual periodicity of Karymsky eruption}
A distinctive feature of 
Karymsky eruptions is its rather robust periodicity which is observed
on two timescales. An explosion, and a following 
quiet 
period, define the first period of activity which
varied between 3 and 20 minutes and sometimes extending up to an hour. 
An explosion itself
can start either abruptly or gradually (Fig.~4) and proceed according to 
one of the following scenarios:
\begin{figure}
\centerline{
\epsfxsize=8cm 
\epsfbox{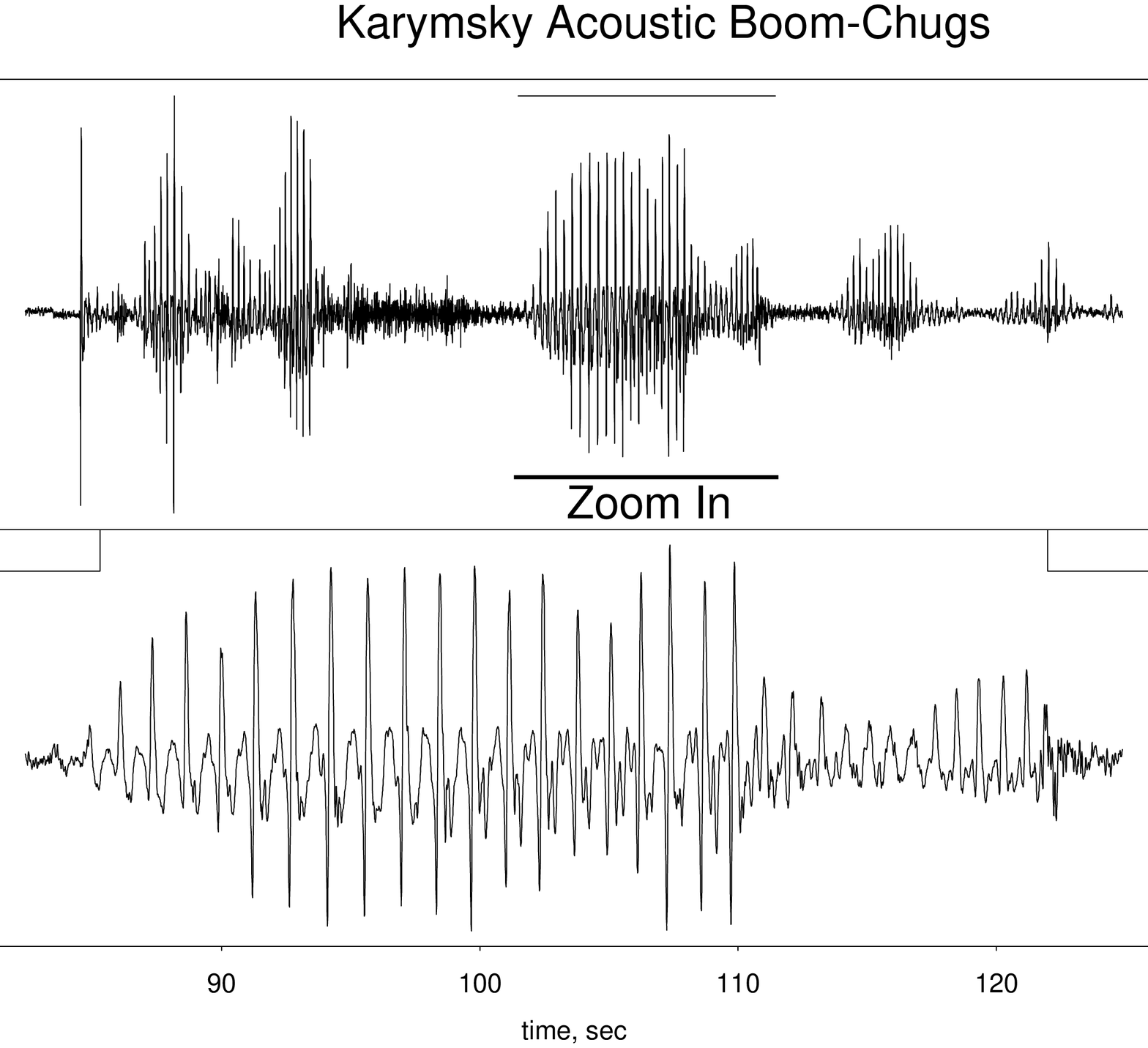}}
\noindent
{\small {\bf Fig.~4.}
Typical acoustic signal of a period of activity of Karymsky volcano
}
\end{figure}
during shorter explosions, the activity decays
quickly and monotonously (in less than 10 seconds), while  the longer 
explosions 
($20-40$ s, rarely 1 minute)
may contain periods of relatively high and low activity.
Often during longer ($>20$s) explosions, the intensity of the acoustic and 
seismic signals emitted by the volcano exhibits chugging, 
i.e. modulations with a typical period of $\sim1$ second. 
The chugging defines the second, shorter, 
period in the cyclic activity of the volcano. 
A typical chugging event consists of 10-20 cycles (Fig.~4). 
The explosion is followed by a quiet period characterized by a 
complete lack of any activity in the crater and usually lasting much longer 
than the explosions themselves. Similar temporal patterns have been also observed
during previous Karymsky eruptions. (Tokarev and Firstov, 1967;
Farberov et al., 1983).

The systematic seismic and acoustic studies of Karymsky volcano were
conducted in 1997-1999 by 3 Russian-US expeditions. The seismograms
presented in Fig.~4 were recorded in 1998 by a three-component wide-band
seismometers CMG 40-T, but the most detailed information about a fine 
structure of the chugging signal
was obtained using a set of infrasound microphones. 
For a more complete description of the seismic and infrasound recordings
at Karymsky the reader is referred to (Johnson et al., 1998;
Johnson, 2000; Johnson and Lees, 2000). 

\section{Existing models of eruption periodicity}
Currently there is no single well-established view on the mechanism causing 
periodic patterns in the 
eruptions of
Strombolian type. Below we mention several possible explanations of 
origins of the periodicity. Lees and Bolton (1998) compared the 
eruption 
process to the steam exhaust from a hydrodynamic system with
a nonlinearly controlled exhaust valve resembling 
a pressure cooker. Manga (1996) 
suggested that the periodicity is caused by a segregation of ascending 
bubbles into 
waves, as in a pint of Guiness beer. Julian (1994)
proposed a 
mechanism of nonlinear hydrodynamic oscillations in the ascending magma flow.
Jaupart and Vergniolle (1989) suggested that oscillations are caused 
by periodical 
collapse of a bubble foam trapped in pockets of the magma chamber.
Recently, there appeared a paper by Delinger and Hoblit (1999)
where an {\it ad hoc} hysteresis model was developed to describe
a periodic behavior of silicic volcanos.
However, self-consistent explanation for
both long- and short-time periodicity 
and extended calm periods in
the activity of Strombolian-type  volcanoes is still missing. 
Our  model, based on viscoelastic
hydrodynamical description of the motion of the magma column with shear 
stress-dependent boundary conditions, 
accounts for both timescales.

\section{The model}

Let us first qualitatively discuss 
the motion of magma in the conduit. Since the temperature
and pressure in the  magma column decrease with the height, 
the viscosity of magma increases rapidly in the 
upper part of the 
conduit. Inspecting eruption products ejected to the surface, 
it is natural to deduce that at the very top of the column the magma is 
almost solid. However, deeper parts of the magma column are still hot 
and much less viscous.
Given this significant difference in rheological properties 
of the upper and lower sections of the magma column, we consider  
the moving magma as consisting of two distinct parts: long lower part 
(LP) and short 
upper part (UP).
We assume that the viscosity of the magma in the LP is negligible. 
Due to the large 
amount of dissolved gases, magma in the LP is compressible. 
We also assume that there is a constant supply of fresh magma to the 
bottom of the LP.
Unlike the LP, magma in the UP is cold and
viscous; in addition to that, as any media near the liquid-solid transition 
point, 
it possesses a certain degree of elasticity. Because of the small length 
of the UP compared to the LP ($\sim 100$~m vs. $\sim 5$~km),
we can neglect compressibility of the UP.
In summary, we consider 
a long cylindrical tube filled with a non-viscous compressible media, 
fed to the bottom with a constant velocity and  
a short viscoelastic plug (UP) on the top. 
The boundary conditions for the moving
UP are controlled by a local value of a shear stress in the 
vicinity of the cylinder wall. When the boundary shear stress is low, 
the plug 
``sticks''
to the cylinder wall and the boundary velocity is zero. As the shear stress 
increases,
the magma near the wall undergoes phase transition (melts) to 
a much less viscous liquid 
state,
and the boundary layer of the plug slips along the walls of the channel.
This shear-induced phase transition happens
when the viscoelastic state with the non-zero shear modulus becomes 
thermodynamically less stable than 
a pure liquid state with zero shear modulus.
Phenomenologically, the transition between stick and slip boundary 
conditions for the
very viscous body is similar to the effect of dry friction, 
where a motion
begins only  
when a driving force exceeds some
threshold value (Persson, 1999, Ch.~8).

Qualitatively, a cycle of the system evolution can be described as follows:
when the pressure in the compressible LP of 
the column 
is low, the shear stress 
$\s$ in the plug is much smaller than the threshold phase 
transition value $\sl$, and the velocity of the UP is much less than
the feeding velocity $V_0$. The difference in feeding and plug velocities 
results in the growth of the pressure of 
the compressible 
magma in LP which, in its turn, increases the shear stress in the plug. 
This stage corresponds to 
the
dormant period between explosions. When the shear stress in the 
boundary layer 
of the UP exceeds the critical value $\sl$, 
the stick-slip phase transition occurs and 
the plug begins to move. This corresponds to the appearance of 
visible signs of 
a volcanic
explosion.

Depending on the parameters of the model, 
the further evolution of the system proceeds 
according to one of the following two scenarios.
For certain conditions, a relaxation of the accumulated in the UP
viscoelastic deformations can result in a low, or even reverse, velocity 
of the 
boundary layers of the UP relative to the walls of the conduit. 
If these conditions persist for sufficient time, 
they can cause a drop of the shear stress below
the slip-stick transition critical value $\st$, which in turn can give rise to
the reverse, slip-stick transition.
However, the stick state will be very short-lived, since the accumulated 
excessive 
pressure in the LP and oscillatory motion of UP 
will result in the immediate increase of the boundary 
shear 
stress and a new stick-slip transition. Length of a period of such
coupled viscoelastic-stick-slip process is controlled by the
density, the elastic modulus, and the characteristic time of the phase 
transition.
In the second scenario, sticking does not occur during short-periodic 
viscoelastic oscillations of the UP, and the period of these oscillations
is roughly equal to that
of a freely oscillating elastic membrane with mixed boundary conditions.
Either of these oscillation processes 
can cause 
the short-period modulation of the activity of the volcano.

During the active phase, the velocity of the  UP is much 
higher than 
the feeding velocity $V_0$ which results in an expansion of the LP and a 
decrease of 
pressure
under the UP.
When the pressure under the 
UP drops so that the shear stress in the UP is
below the slip-stick threshold value $\st$, a 
``long-term'' sticking of the plug occurs, 
and the system enters 
a new quiet period of evolution.
Because of the hysteresis ($\sl<\st$) associated with 
the first order nature of the stick-slip transition,
it takes some time to build enough pressure for a new explosion to begin.
This cycle corresponds to the second, longer period 
(10-30 minutes) in the dynamics of 
eruption.

Although simple in nature, this model  qualitatively explains both scales of 
periodicity of  
Karymsky volcano and does not contradict to the observations.
Below a formal analysis of the dynamics of the model is presented.

Let us consider a cylinder (LP) of the radius $R$ and the length $L$, 
filled with 
an ideal non-viscous compressible medium feeded from the bottom
with a constant velocity $V_0$. The top of LP ends with a short 
viscoelastic 
plug (UP) of length $l$, $l\ll L$ (Fig.~5). 
\begin{figure}
\centerline{\epsfxsize=2cm \epsfbox{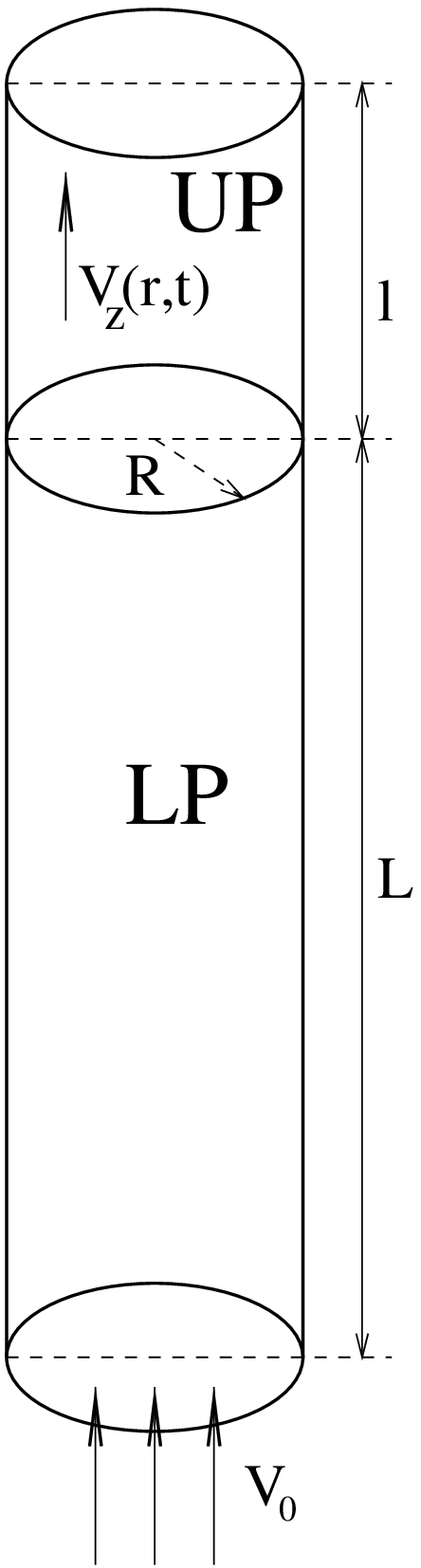}}
\noindent
{\small {\bf Fig.~5.}
Sketch of the magma conduit.
}
\end{figure}
The velocity field in the UP, of  density $\r$ and 
driven by pressure $P$
and stress tensor $\S_{ij}$, is described by the Navier-Stokes
equation,
\begin{equation}
\label{ns}
\r {\partial v_j \over \partial t}=-{\partial P\over \partial x_j}
+{\partial \S_{ij} \over \partial x_i}.
\end{equation}
Because of the high viscosity and the low velocity of magma in the UP,
the nonlinear (convective) term is omitted.
To account for the viscoelastic properties of the UP 
we use the simple 
linear Maxwell model (see, for example, Shore et al.),  with a characteristic 
memory time $\t_m$,
\begin{equation}
\label{m}
\t_m {\partial \S_{ij} \over \partial t}=-\S_{ij}
+\n e_{ij}.
\end{equation}
Here $\n$ is the dynamic viscosity, and $e_{ij}\equiv \partial v_i/
\partial x_j+\partial v_j/ \partial x_i$ is the shear rate tensor.
Incompressibility condition for the magma in the UP reads:
\begin{equation}
\label{uc}
{\partial v_i \over
\partial x_i}=0.
\end{equation}

For simplicity, we assume that the motion of the 
plug is axially 
symmetric and
select the $Z$ direction of our cylindrical coordinate system parallel the axis
of the cylinder directed along the magma flow. Also, we also consider the 
pressure 
gradient constant through
the UP and being equal to ${- P_0\over l}$, where $P_0$ is a pressure at the 
bottom of the UP.
Using (\ref{m},\ref{uc}), the Navier-Stokes equation (\ref{ns}) 
for $V_z$ ($Z$-component of the velocity of 
the plug averaged over 
the plug length) can be transformed to:
\begin{equation}
\label{v}
\r {\partial V_z (r,t)\over \partial t}={P_0(t)\over l}+{\n \over \t_m}
\int^t
\exp\left (-{t-t' \over \t_m}\right ) \D_r V_z(r,t')dt'.
\end{equation}
Here $\D_r \equiv
\partial^2/\partial r^2+ 1/r\; \partial/\partial r$ is the radial part of the 
Laplacian operator
in the cylindrical coordinates.

From the conservation of  
mass of the
compressible media in the LP, 
it follows for the pressure $P_0$ on the bottom surface 
of the UP,
\begin {equation}
\label {p}
V(t)-V_0=-\beta L {d P_0 (t) \over dt},\: V(t)\equiv 2\int_0^R V_z(r,t)r d r/R^2,
\end {equation}
where $\beta\equiv -1/v (\partial v /\partial p)_T$ is the compressibility 
(isothermal) of the
media in the LP, and $V(t)$ 
is the velocity of the UP $V_z(r,t)$,
averaged over the perpendicular crossection. 

To complete the description of the dynamics of the system, the equations
(\ref{v},\ref{p}) must be complemented by boundary conditions for the 
velocity on the conduit walls, $V_z(r=R,t)$. 
These boundary conditions are 
controlled by the kinetics of the first-order phase transition driven by the
value of the shear stress $\sb(t) \equiv \s_{rz}(t, R)$ near the wall
and are written in a general mixed form:
\begin{equation}
\label{bc}
V_z (R,t)=-R\p(t)\left. {\partial V_z(r,t)\over \partial r}\right|_{r=R}.
\end{equation}
The quantity $R\p(t)$ is usually called a ``slipping length''. 
It shows at
what distance inside the wall the linearly extrapolated velocity becomes equal 
to zero. 
For $\p(t)=0$ the boundary conditions
for the velocity are ``stick'', i.e. $V_z (R,t)=0$; for $\p(t)> 0$ the 
boundary layers of the UP ``slip'' along the walls of the cylinder
with some finite velocity, $V_z (R,t)>0$. 

The time evolution of the dimensionless slipping length $\p(t)$
is determined from the time-dependent Ginzburg-Landau equation,
which describes relaxation of a system with a first-order phase transition:
\begin{equation}
\label{gl}
\t_{GL}{d\p \over dt}=-{d F\over d \p},
\end{equation}
where $\t_{GL}$ sets a phase transition timescale.
The free energy $F(\p)$ is a usual double-well potential  
with a tilt depending on the value of the shear stress $\sb(t)$ on 
the boundary of the UP. We choose a double-parabola potential which has
fixed positions of both minima, $\p=0$ and $\p=\p_0$:
\begin{equation}
\label{f}
F(\p)= 
   \left\{ 
\begin{array}{ccc}
    \p^2,& \p < \p' \\ 
   (\p-\p_0)^2-h,& \p \geq \p' 
\end{array} \right. 
\end {equation}
Here $h$ is a tilt parameter; matching point $\p'$ is determined from 
continuity requirement, $\p'=(\p_0^2-h)/2\p_0$.
To determine $h$ and to illustrate how Eqs.~(\ref{gl},\ref{f}) work,
let us consider a typical stick-slip cycle. 
We go around a hysteresis curve shown in Fig.~6.
\begin{figure}
\centerline{\epsfxsize=8cm \epsfbox{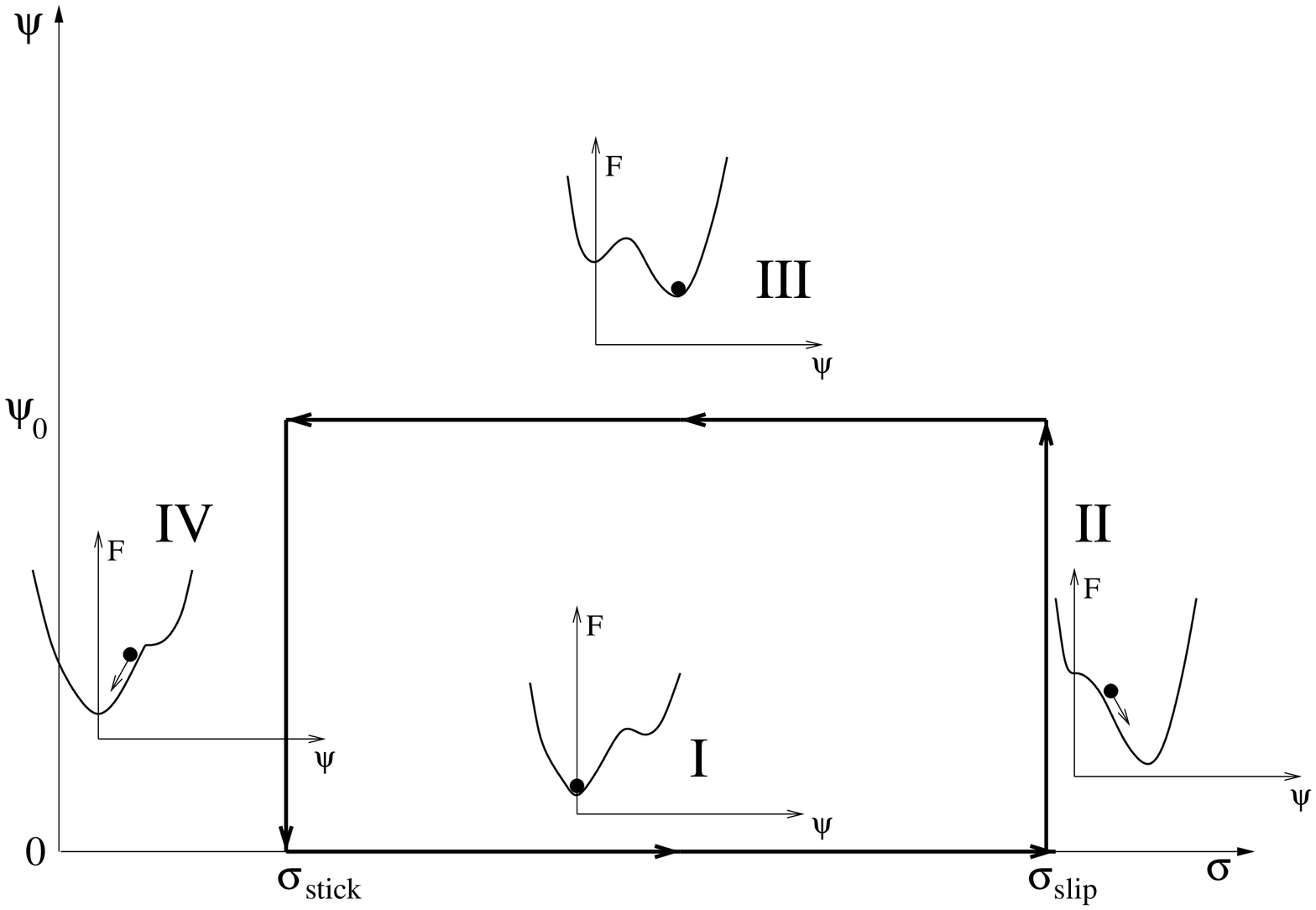}}
\noindent
{\small {\bf Fig.~6.}
Hysteresis curve {\bf I-II-III-IV} for the stick-slip and the slip-stick 
transitions.
Inserts show typical sketches of $F(\p)$ for all branches of the curve.}
\end{figure}

When the shear stress $\sb(t)$ is small and growing (line {\bf I}),
the boundary condition is ``stick'' and stationary,
$\p={d F\over d \p}=0$.
As the shear stress exceeds the critical value $\sl$, $\p(t)$ starts
moving towards $\p_0$ with the rate $\sim 2\p_0/\t_{GL}$ (line {\bf II}). 
After
$p$ reaches $\p_0$, the boundary condition becomes ``slip''
and stationary again, and
the shear stress starts decaying  (line {\bf III}).
After $\sb(t)$ drops
below $\st$, $\p$ relaxes to 0 (line {\bf IV}).
The state {\bf I} corresponds to the dormant period, the state {\bf III} 
corresponds
to the explosion without intermediate sticking; the transition intervals 
{\bf II, IV} 
are usually very short.   
The free energy tilt parameter $h$ is linearly connected to $\sb$:
\begin{equation}
\label{h}
h=\p_0^2 \left [ {2\sb \over \sl-\st}-{\sl+\st \over \sl- \st} \right]. 
\end {equation}

Similarly to the Eq.~(\ref{v}), in the framework of the Maxwell model 
the shear stress 
$\sb(t)$ is expressed as a convolution of a shear rate 
${\partial V_z(r,t')\over \partial r}$ with an exponential memory function
$\exp(-{t-t' \over \t_m})$
\begin{equation}
\label{sh}
\sb(t)=
{\n \over \t_m}\int^t
\exp(-{t-t' \over \t_m}) \left. {\partial V_z(r,t')\over \partial r}
\right |_{r=R}\; dt'
\end{equation}

The equations (\ref{v}--\ref{sh}) completely define the dynamics 
of our model.
A very similar set of equation was derived by Shore et al. (1997) for the physically
equivalent model of polymer extrusion. 
Because of the strong non-linearity of the Eq.~(\ref{gl}), an analytic
solution of these equations is impossible. 
We refer readers to Shore et al., 1997, for a detailed description of a
linear stability analysis of this system.
Contrary to a stable flow, a dynamical regime corresponding to volcanic 
activity is  strongly unstable and nonlinear. However, large separation
of characteristic timescales allows us to give reasonable
analytical estimates of important temporal features of the model.

Let us first look at the shortest timescale which corresponds to
small oscillations of 
the UP when it slips along the conduit walls without intermediate sticking.
For this
timescale the pressure term in (\ref{v}) is considered to be constant. 
Dynamics of the UP is physically equivalent
to the oscillation of an elastic circle membrane with the mixed boundary conditions
(\ref{bc}). Only the lowest harmonic is considered,
\begin{equation}
\label{vj}
V_z(r,t)=V_0 e^{-i\om t}J_0 \left({r\over R} \l \right),
\end{equation}
where $\l$, a number of order of one, depends on the slipping length $\p$:
$J_0(\l)=J_1(\l)\l \p$. Here $J_0$ and $J_1$ are the zero- and first-order
Bessel functions. After plugging Eq.~(\ref{vj}) into  Eq.~(\ref{v}) and 
disregarding the constant pressure term, we obtain the following dispersion 
relation:
\begin{equation}
\label{om}
\om={-i\pm \sqrt{4{\l^2 \n \t_m\over R^2 \r}-1}\over 2 \t_m}.
\end{equation}
This yields for a period $T_{osc}$ of not too overdamped UP oscillations 
($T_{osc} \gg \t_m$):
\begin{equation}
\label{omp}
T_{osc}\approx{2 \pi R \over \l}\sqrt{\r\t_m\over \n }.
\end{equation}
Taking into account a commonly used relation between the Maxwell time
and shear modulus $G$,
$\t_m\approx \n/G$ (see, for example, Webb (1997)), we observe that
Eq.~(\ref{omp}) reduces to a usual expressions for a period of a small 
free oscillations of an elastic membrane, 
\begin{equation}
\label{T}
T_{osc}\approx{2 \pi R \over \l}\sqrt{\r\over G }.
\end{equation}

Now we go to much longer timescales, $T\gg \{T_{osc}, \t_m\}$ and look first 
at the steady state UP boundary shear stress $\sb=\left. 
{\partial V_z(r)\over \partial r}
\right |_{r=R}$. Since this timescale is much greater than the elastic 
memory time $\t_m$, the viscoelastic term in (\ref{v}) is relaxed to
a usual viscous term, $\n \D_r V_z(r,t)$. For a steady state,
the averaged over the crossection of the conduit velocity
$V(t)$ must be equal to 
$V_0$.
For an arbitrary slipping length $\p$, 
a straightforward calculation yields:
\begin{equation}
\label{si}
\sb={4 V_0\n \over R} {1\over 1+4\p}.
\end{equation}
It immediately follows that in order for stick-slip
repeating cycles to happen, the ``stick'' steady state shear stress ($\sb$ for
$\p=0$) should be
greater than $\sl$ and ``slip'' steady state shear stress ($\sb$ for
$\p=\p_0$)
should be less than $\st$. These conditions imply for the feeding velocity 
$V_0$:
\begin{equation}
\label{vosc}
{\sl R\over 4} < V_0 < {\st R (1+4\p_0) \over 4}.
\end{equation} 

Besides the steady state properties, we can also evaluate characteristic times
between the stick-slip and the slip-stick phase transitions, i.e. duration
of the dormant and explosive periods. Assuming that these times are much longer
than both $T_{osc}$ and $\t_m$, we again 
disregard all memory effects in the viscous term
in Eq.~(\ref{v}) and consider the flow inertialess by omitting the 
left-hand-side
term ${\partial V_z (r,t)\over \partial t}$. After averaging the remaining part
of Eq.~(\ref{v}) over the crossection of the flow and plugging it in (\ref{p}),
we obtain for arbitrary slipping length $\p$:
\begin{equation}
\label{vl}
\int^t{V_o-V(t')\over \be L l}dt'-{8\n V(t)\over R^2(1+4\p)}=0.
\end{equation} 
Combining the solution of this equation,
\begin{equation}
\label{vls}
V(t)=V_o+C e^{-{t\over t_0}},\;\;t_0\equiv {8\n \be L l \over R^2 (1+4\p)},
\end{equation} 
with expressions for stick and slip
velocity that follow from (\ref{si}), we obtain for dormant 
($T_d$) and explosive 
($T_e$) times:
\begin{equation}
\label{td}
T_d={8\n \be L l\over R^2}\ln{4 \n V_0-{\st R}\over 4 \n V_0-{\sl R}} 
\end{equation} 
\begin{equation}
\label{te}
T_e={8\n \be L l\over R^2(1+4\p_0)}\ln{{\sl R(1+4\p_0)}-4 \n V_0\over 
{\st R(1+4\p_0)}-4 \n V_0}. 
\end{equation} 

\section{NUMERICAL RESULTS}

To check our analytical predictions and get an overall
view of the system dynamics, we solved the Eqs.~(\ref{v}--\ref{sh})
numerically.
We used a simple Eulerian 
finite-difference scheme with a uniform radial grid of 50-100 points. 
All the integrals are approximated by trapezoid formula.
Given that time step is sufficiently small to avoid von Neumann-type 
instabilities, this simplest difference scheme proved to be sufficiently 
robust.

We chose the numerical parameters to be in general accordance
with the literature (Murase and McBirney, 1973;, 
 Borgia and Linneman, 1990;
Griffiths and Fink, 1993;, Bagdassarov and Dorfman, 1994; Webb, 1997;
Bagdassarov and Dorfman, 1998)
and with our estimates 
(\ref{T},\ref{vosc},\ref{td},\ref{te}). The following values were used:
\begin{list}{$$}{\itemsep=0pt \leftmargin=0cm}
\item Density $\r=2400\:kg/m^3$
\item Dynamic viscosity $\n=10^9\:Pa\,s$
\item Conduit raidus $R=50\:m$
\item UP length $l=100\:m$
\item LP length $L=5\:km$
\item Compressibility $\be=4\,10^{-10}\:Pa^{-1}$
\item Maxwell time $\t_m=0.4\:s$
\item Ginzburg-Landau time $\t_{GL}=10^{-3}\: s$
\item Average velocity $V_0=2\times10^{-3}\: m/s$
\item Slipping shear stress $\sl=10^5\:Pa$
\item Sticking shear stress $\st=3\,10^4\: Pa$
\item Dimensionless slipping length $\p=10$
\end {list}

With this choice of numerical parameters a no-stick explosion
takes place, during which the UP moves along the walls of the conduit
with the constant slipping length $\p R$ until it finally sticks
to the wall at the end of the explosion. 
For such explosions a choice of phase 
transition time $\t_{GL}$ does not affect the dynamics, given that
it is shorter than explosion time, $\t_{GL}\ll T_e$.

Plots of averaged over crossection velocity $V(t)$ and pressure gradient
vs. time
are presented in Fig.~7. 
The simulation results for dormant and explosion times, as well for
short-term oscillation period, $T_d^s\approx 530$ s, $T_e^s\approx 21$s,
$T_{osc}^s\approx 0.69$ s are in good agreement with our simple analytical 
estimates, $T_d=497$ s, $T_e=20$ s, and $T_{osc}=0.69$ s.
To calculate $T_{osc}$ we have to solve the equation $J_0(\l)=J_1(\l)\l \p$, which
for $\p=10$ yields $\l \approx 0.43$. For the stick boundary conditions 
($\p=0$), 
$\l\approx 2.3$; which gives for the period of oscillations 
at the end of explosion cycle when UP sticks to the walls 
$T_{osc}'\approx 0.13$,
which is again in very good agreement with the simulation data.
\begin{figure}
\centerline{\epsfxsize=8cm \epsfbox{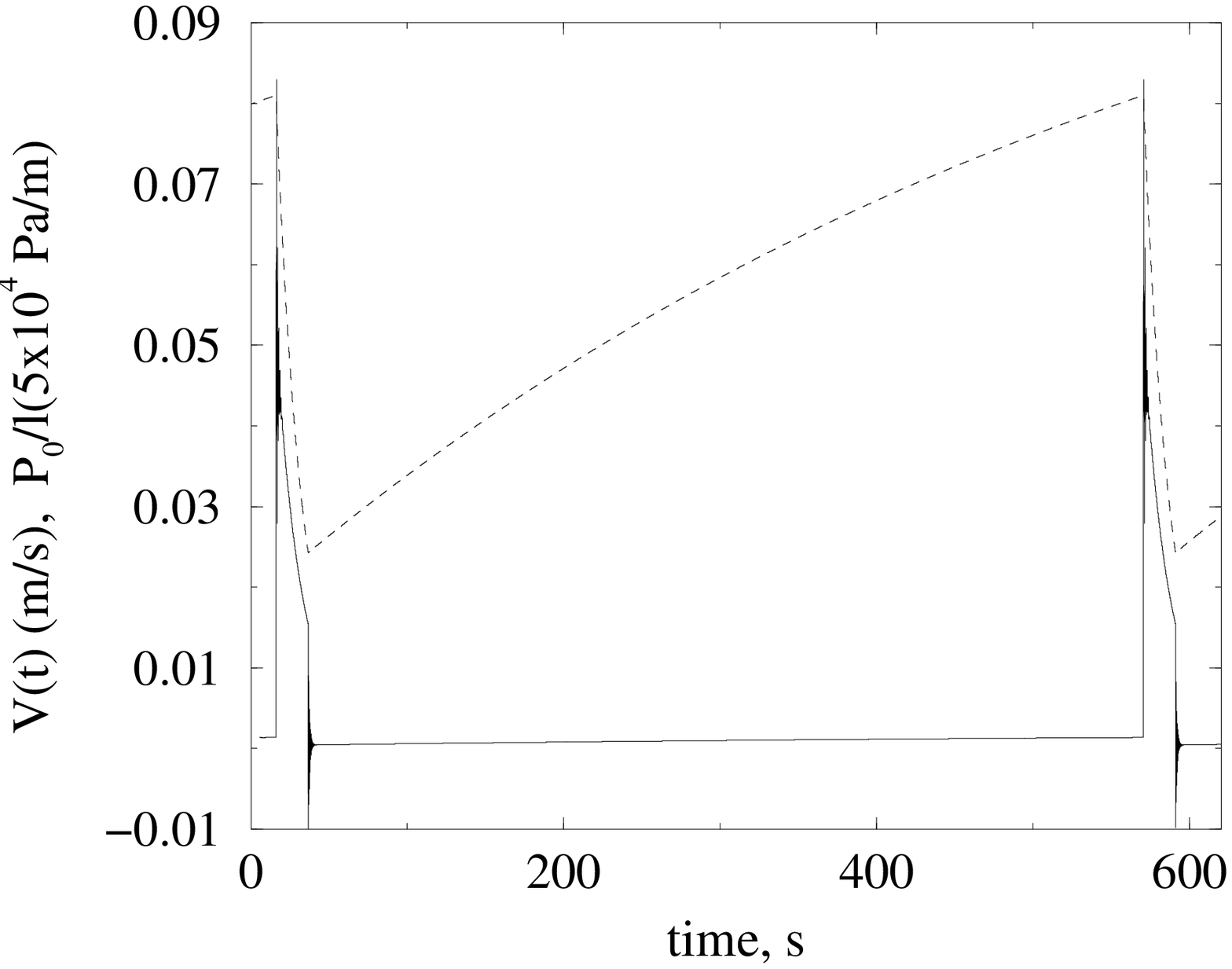}}
\centerline{\epsfxsize=8cm \epsfbox{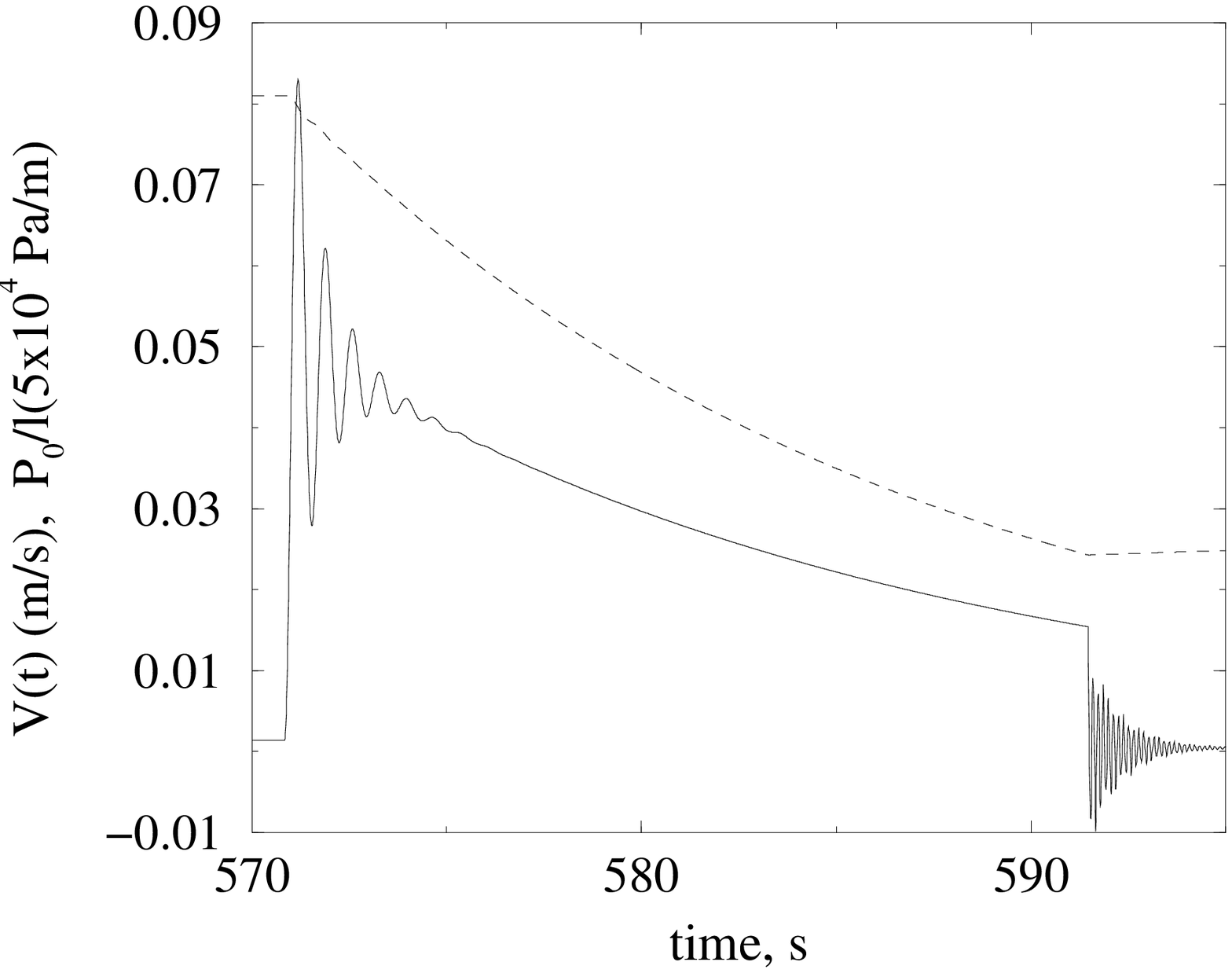}}
\noindent
{\small {\bf Fig.~7}. 
Time dependence of the spatially averaged UP velocity (solid line) 
and the rescaled pressure gradient (in $5\times 10^4$ Pa/m, dashed line).}
\end{figure}
Another possible scenario for an eruption is when during explosion cycle
the UP sticks to the walls numerous times. It happenes when the
shear stress at the wall, modulated by the free membrane-like 
oscillations of the UP, drops below the sticking value $\st$, and the
slip-stick phase transition is sufficiently fast, $\t_{GL}\ll T_{osc}$.
To simulate this regime, we increase the value of $\st$ from
$3\,10^4$ to $8.5\,10^4$ Pa and keep all other parameters constant.
The long-time dynamics, characterised by the times $T_e$ and $T_d$ only
slightly changes since both of these times depend on $\st$ logarithmically.
However, a short-time dynamics, previously described by decaying free
oscillations with a period $T_{osc}$, changes radically and becomes chaotic 
(Fig.~8).
\begin{figure}
\centerline{\epsfxsize=8cm \epsfbox{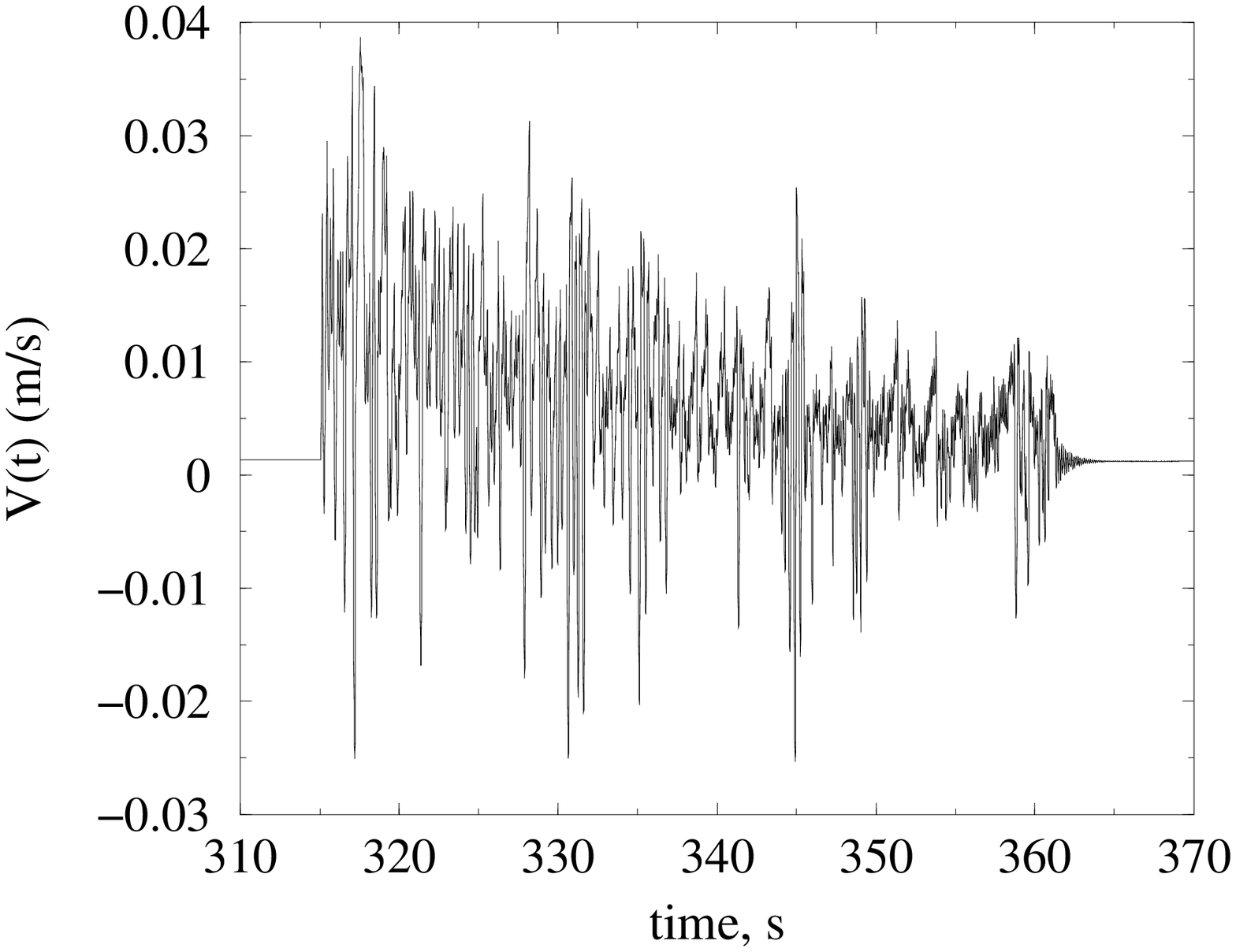}}
\centerline{\epsfxsize=8cm \epsfbox{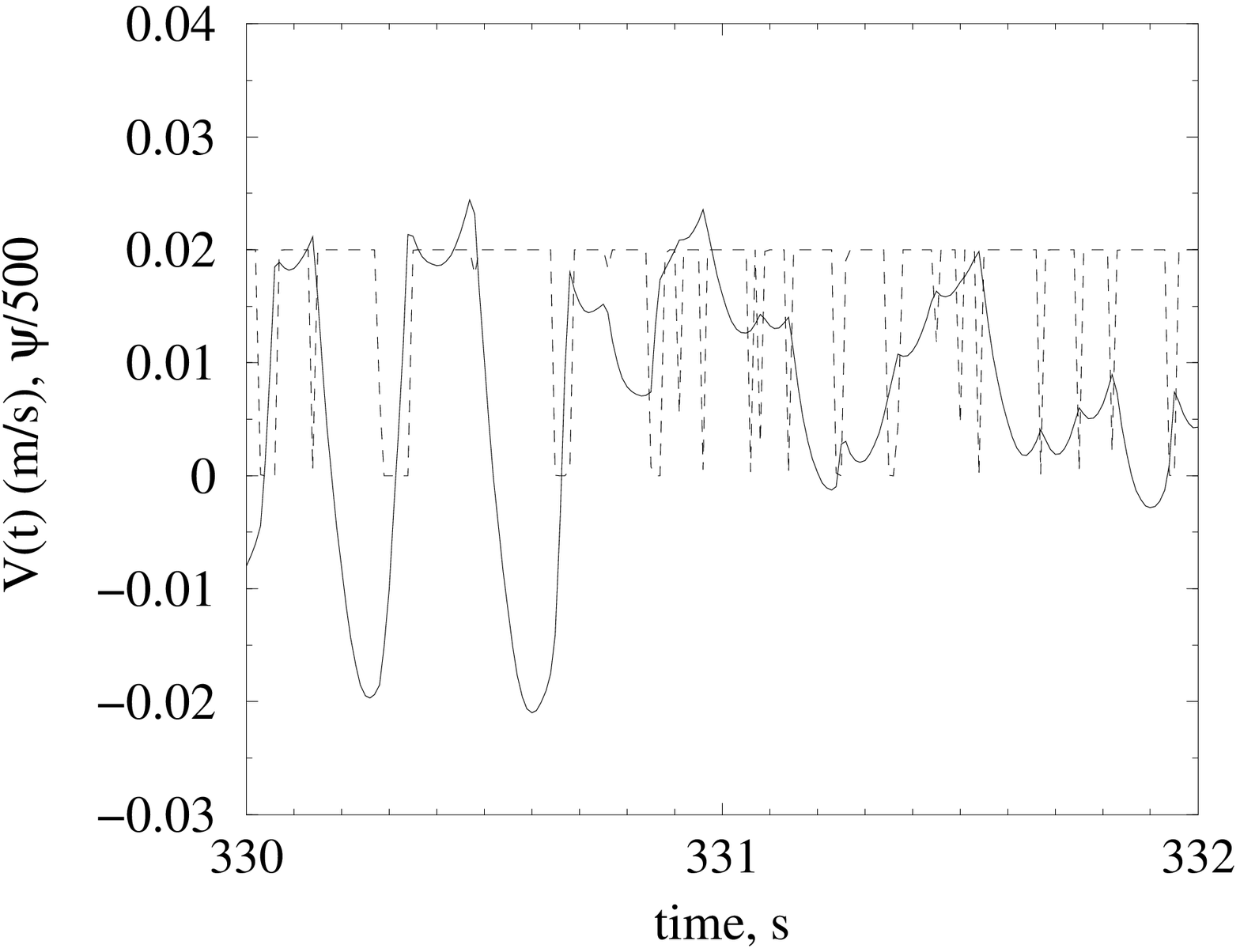}}
\noindent
{\small {\bf Fig.~8}. 
Time dependence of the spatially averaged UP velocity $V(t)$ (solid line) 
and the rescaled dimensionless slipping length $\p(t)/500$ (dashed line).}
\end{figure}
A typical time separation between consecutive velocity maxima ($\approx 0.3$ 
s for the first three maxima in the bottom part of Fig.~8)
becomes somewhat 
smaller than free oscillation period $T_{osc}\approx 0.7$. 
Qualitatively, it happenes when during an oscillation
the velocity increases so that shear stress drops below $\st$, a sticking occurs
that results in a rapid decrease of velocity, so the oscillation
is ``truncated'' before it reaches its maximum.
Another distinct feature of this regime is that the short-time chaotic 
modulation of the
UP velocity continues through the whole explosion phase, while in 
the non-sticking regime, the oscilations decay after a few periods.

\section{Discussion}

In the previous sections the long-term behavior and 
two possible scenarios of the short-time activity of the model
based on viscoelastic hydrodynamics and stick-slip transition were described.
Patterns in the time dependence of the UP velocity, presented 
in Figs.~7,8, qualitatively
match the seismic and acoustic signals (Fig.~4) recorded
at Karymsky volcano. Yet nothing has been said about a 
mechanism describing generation of seismic and acoustic waves themselves.
The oscillating UP and related pressure oscillations in the LP
can cause seismic waves either directly or act as a ``pacemaker'' 
for other processes. These processes can include a stimulated emission 
of gases either
through creation and change of geometry of cracks in the UP or via 
degassing of magma under action of standing and running compression waves.
Since these mechanisms have a strongly nonlinear character, their coupling
to the UP oscillation can modify the oscillation period and decay patterns.
We leave consideration of these phenomena for future study.

\section{conclusion}

As a result of the visual, seismic, and acoustic observations at Karymsky 
volcano,
two scales of temporal periodicity were observed in the eruptive dynamics. 
First, the repetitive explosions which
produced tha ash and gas clouds and volcanic bombs in 
10-20 minutes 
intervals, and secondly, the periodic modulation 
of eruptive activity (chugging) with a typical timescale of order of 1 s.
The existence of these two scales of temporal periodicity are not limited to the
1996-2000 eruption of Karymsky volcano, similar timescales were observed
during previous Karymsky eruptions and on other andesite volcanos in the world.

A model of the motion of the magma column was suggested that accounts for both 
timescales in the eruption of Karymsky and other similar andesite 
Strombolian-type volcanos.
Low- and high-frequency periodicity is caused by two distinct ways of 
accumulation of the elastic energy. The viscoelastic properties of the UP 
causes the high frequency tremor, while the  
compressibility of  magma in the LP 
explains the low frequency cycles. The shear-stress-dependent 
stick-slip phase transition in the UP introduces
the hysteresis into the dynamics of the magma column motion and explains the unharmonicity
of the oscillations and long dormant periods in the eruption dynamics
Physical simplicity of the model allows analytic estimates for
the long and short periods of the explosive activity to be obtained.

\section{Acknowledgments}
The authors are grateful to Martin Grant for stimulating discussions on 
the polymer extrusion problem, to Julie M. Kneller 
for carefully reading the manuscript, and especially to
Emily Brodsky for providing
important references, numerical values for magma parameters, 
and carefully reading the manuscript. 
This work was partially supported by the U.S. National Science Foundation.

\section{Bibliography}
\begin{list}{$$}{\itemsep=0pt \leftmargin=0cm}
\item         Bagdassarov~N.~S., Dingwell~D.~B. and Webb~S.~L., 1994.
		   {\it Viscoelasticity of crystal- and bubble-bearing rhyolite
                   melts}, Phys. Earth Planet. Inter., 83: 83-99.

\item         Bagdassarov~N. and A.~Dorfman~A., 1998. {\it Viscoelastic behavior of
                   partially molten granites}, Tectonophysics, 290: 27-45.

\item         Benoit~J.~P., McNutt~S.~R., 1997. {\it New constraints on source of
			volcanic tremor at Arsenal Volcano, Costa Rica, using broadband
			seismic data}. Geoph. Res. Lett., 24: 449-452.

\item         Bird~R.~B., Armstrong~R.~C., and Hassager~O., 1987. {\it
                   Dynamics of Polymeric Liquids}. Vol.1: Fluid mechanics,
                   Wiley, New York.

\item	      Borgia~A. and Linneman~S.~R., 1990. 
                   {\it On the Mechanism of Lava Flow 	
			Emplacement and Volano Growth:
		       Arenal, Costa Rica}. Lava Flows and Domes, ed. 
			J.H. Fink, Springer-Verlag.

\item         Delinger~R.~P. and Hoblitt~R.~P., 1999. {\it Cyclic eruptive 
                   behavior of silicic volcanoes}. Geology, 27:
                   459-462.

\item	      Farberov~A.~I., Strocheus~A.~B., and Pribulov~E.~S., 1983.
			{\it Studies of weak seismicity of Karymsky volcano, 
			August 1978}. Vulkanologija i Seismologija, 3: 78-89
			(in Russian).

\item	      Griffths~R.~W. and J.~H.~Fink, 1993.
		                                {\it Effects of surface
						cooling on the spreading of
						lava flows and domes}. J. Fluid
						Mech., 252: 667-702.

\item	     Hess~P.~S., 1989. {\it Origins of
						Igneous Rocks}, Harvard Univ. Press.,
						Cambridge, USA.

\item         Jaupart~C. and Vergniolle~S., 1989. 
			{\it The generation and collapse of a foam
                   layer at the roof of a basaltic magma chamber}. J. Fluid Mech.,
                   203: 347-380. 

\item         Johnson~J.~B., 2000. {\it Interpretation of infrasound generated by 
			erupting volcanoes and seismological energy partitioning 
			during Strombolian Explosions}. PhD Thesis, Univ. of Washington,
                        Seattle.

\item         Johnson~J.~B. and Lees~J.~M., 2000.
                   {\it Plugs and Chugs - Strombolian activity at Karymsky, Russia, 
                   and Sangay, Ecuador}. J. Volc. Geotherm. Res., 101: 67-82.
		   
\item	      Johnson~J.~B., Lees~J.~M., and Gordeev~E.~I., 1998. {\it Degassing 
                            explosions at Karymsky volcano, Kamchatka}. 
				Geophys. Res. Lett., 25: 3999-4000.

\item       Julian~B.~R., 1994. {\it Volcanic tremor: Nonlinear excitation by fluid flow}.
                   J.\ Geoph.\ Res., 99: 11859-11877.

\item       Lees~J.~M. and Bolton~E.~M., 1998. {\it Pressure cookers
                   as volcano analogues}. EOS, Trans. Am. Geoph. Un.,
                   79 (45), Fall Meeting Supp., F620.

\item       Magus'kin~M.~A., Enman~V.~B., Seleznev~B.~V., and Shkred~V.~I., 1982.
			{\it Peculiarities of the Earth core displacement on Karymsky
			volcano in 1970-1981 from geodesical and photogrammetric data}.
			Vulkanologija i Seismologija, 4: 49-64 (in Russian).

\item       Manga~M. 1996. {\it Waves of bubbles in basaltic magmas and lavas}.
                   J.\ Geoph.\ Res., 101: 17457-17465.

\item       Murase~T. and McBirney~A.~R., 1973. {\it Properties of some
                   common igneous rocks and their
                   melts at high temperatures},
                   Geol.\ Soc.\ Am.\ Bull., 84: 3563-3592.

\item       Persson~B.~N.~J.~ , 1999. {\it Sliding Friction}. Springer-Verlag, Ch.~8.

\item      Shirokov~V.~A., Ivanov~V.~V., Stepanov~V.~V., 1988. {\it
			On the deep structure of Karymksy volcano and peculiarities
			of its seismicity: local seismic network data}.
			Vulkanologija i Seismologija, 3: 71-80 (in Russian).

\item       Shore~J.~D.~, Ronis~D., Piche~L., and Grant~M., 1997. {\it Theory of melt 
                   instabilities in the capillary flow of polymer melts}.
                   Phys.\ Rev.\ E, 55: 2976-2992.

\item       Tokarev~P.~I. and Firstov~P.~P., 1967. {\it Seismological studies of
			Karymsky volcano}. Bulluten' Vulkanologicheskih Stanciy,
			43: 9-22 (in Russian).

\item       Webb~S., 1997. {\it Silicate melts: Relaxation, Rheology,
                   and the Glass transition}. Rev.\ of\ Geophys., 35:
                   191-218.

\item       Zubin~M.~I, Ivanov~B.~V., Shteinberg~G.~S., 1971. 
			{\it Structure of Karymsky
			volcano, Kamchatka, and some aspects of caldera genesis}.
			Geologiya i Geofizika, 1: 73-81 (in Russian).

\end{list}

\end{document}